\documentclass[12pt,psfig]{article}
\usepackage{amssymb}
\usepackage{amsmath}
\usepackage{epsfig}
\makeatletter
\def\@cite#1#2{{[{#1}]\if@tempswa\typeout
{IJCGA warning: optional citation argument
ignored: `#2'} \fi}}

\newcount\@tempcntc
\def\@citex[#1]#2{\if@filesw\immediate\write\@auxout{\string\citation{#2}}\fi
  \@tempcnta\z@\@tempcntb\m@ne\def\@citea{}\@cite{\@for\@citeb:=#2\do
    {\@ifundefined
       {b@\@citeb}{\@citeo\@tempcntb\m@ne\@citea\def\@citea{,}{\bf ?}\@warning
       {Citation `\@citeb' on page \thepage \space undefined}}%
    {\setbox\z@\hbox{\global\@tempcntc0\csname b@\@citeb\endcsname\relax}%
     \ifnum\@tempcntc=\z@ \@citeo\@tempcntb\m@ne
       \@citea\def\@citea{,}\hbox{\csname b@\@citeb\endcsname}%
     \else
      \advance\@tempcntb\@ne
      \ifnum\@tempcntb=\@tempcntc
      \else\advance\@tempcntb\m@ne\@citeo
      \@tempcnta\@tempcntc\@tempcntb\@tempcntc\fi\fi}}\@citeo}{#1}}
\def\@citeo{\ifnum\@tempcnta>\@tempcntb\else\@citea\def\@citea{,}%
  \ifnum\@tempcnta=\@tempcntb\the\@tempcnta\else
   {\advance\@tempcnta\@ne\ifnum\@tempcnta=\@tempcntb \else
\def\@citea{--}\fi
    \advance\@tempcnta\m@ne\the\@tempcnta\@citea\the\@tempcntb}\fi\fi}
\makeatother

\newcommand{\gsim}{\lower.7ex\hbox{$\;\stackrel{\textstyle>}{\sim}\;$}}
\newcommand{\lsim}{\lower.7ex\hbox{$\;\stackrel{\textstyle<}{\sim}\;$}}
\newcommand{\be}{\begin{equation}}
\newcommand{\ee}{\end{equation}}
\newcommand{\bea}{\begin{eqnarray}}
\newcommand{\eea}{\end{eqnarray}}

\usepackage{amssymb}

\def\baselinestretch{1}
\begin{document}
\catcode`@=11
\newtoks\@stequation
\def\subequations{\refstepcounter{equation}%
\edef\@savedequation{\the\c@equation}%
  \@stequation=\expandafter{\theequation}
  \edef\@savedtheequation{\the\@stequation}
  \edef\oldtheequation{\theequation}%
  \setcounter{equation}{0}%
  \def\theequation{\oldtheequation\alph{equation}}}
\def\endsubequations{\setcounter{equation}{\@savedequation}%
  \@stequation=\expandafter{\@savedtheequation}%
  \edef\theequation{\the\@stequation}\global\@ignoretrue

\noindent}
\catcode`@=12
\begin{titlepage}
\title{{\bf
Flux Compactifications: Stability \\and Implications for Cosmology}}
\vskip2in
\author{
{\bf Ignacio Navarro$$\footnote{\baselineskip=16pt E-mail: {\tt
ignacio.navarro@durham.ac.uk}}}
$\;\;$and$\;\;$
{\bf Jose Santiago$$\footnote{\baselineskip=16pt E-mail: {\tt
jose.santiago-perez@durham.ac.uk}}}
\hspace{3cm}\\
 $$~{\small IPPP, University of Durham, DH1 3LE Durham, UK}.
}

\date{}
\maketitle
\def\baselinestretch{1.15}
\begin{abstract}
\noindent

We study the dynamics of the size of an extra-dimensional manifold
stabilised by fluxes. Inspecting the potential for the 4D field
associated with this size (the radion), we obtain the conditions
under which it can be stabilised and show that stable
compactifications on hyperbolic manifolds necessarily have a
negative four-dimensional cosmological constant, in contradiction
with experimental observations. Assuming compactification on a
positively curved (spherical) manifold we find that the radion has
a mass of the order of the compactification scale, $M_c$, and
Planck suppressed couplings. We also show that the model becomes
unstable and the extra dimensions decompactify when the
four-dimensional curvature is higher than a maximum value. This in
particular sets an upper bound on the scale of inflation in these
models: $V_\mathrm{max} \sim M_c^2 M_P^2$, independently of
whether the radion or other field is responsible for inflation. We
comment on other possible contributions to the radion potential as
well as finite temperature effects and their impact on the bounds
obtained.

\end{abstract}

\thispagestyle{empty} \vspace{5cm}  \leftline{}

\vskip-22.5cm \rightline{} \rightline{IPPP/04/18}
\rightline{DCPT/04/36} \rightline{hep-th/0405173} \vskip3in

\end{titlepage}
\setcounter{footnote}{0} \setcounter{page}{1}
\newpage
\baselineskip=20pt

\section{Introduction}

In recent years great progress has been made in the measurement of the
parameters that control the evolution of our universe and experiments
like WMAP~\cite{Spergel:2003cb} 
have provided us with precision data to compare with the
predictions of theories of the early universe such as inflation. In
this way a standard cosmological model has emerged, the so-called
Lambda Cold Dark Matter ($\Lambda$CDM) or Concordance
Model~\cite{Peiris:2003ff}. This
model assumes inflation in the early universe but one of its most
surprising features is the small but nonzero observed cosmological
constant at present times, that poses a deep fundamental challenge for
theorists. There is then a great deal of activity in what could be
called cosmological phenomenology, or trying to figure out the
implications for cosmology of the available high energy theories of
fundamental physics. Many of these hypothesised theories, like String
Theory, are formulated in a higher dimensional space, so one commonly
assumes that the extra dimensions are compactified with a small
volume. The purpose of this letter is to extract cosmological
implications of the scenario in which the size of the extra dimensions
is stabilised using fluxes in the higher-dimensional space.

Flux compactifications, although an old topic of research in high
energy physics (see for instance \cite{Freund:1980xh}), have received
a lot of
attention recently \cite{Lukas:1996iq}. In
the next section we work out in detail the compactification of $n$
extra dimensions in a manifold of constant curvature (hyperbolic
or spherical, depending on the sign of the curvature) using
fluxes. We will rephrase the dynamics of the size of the extra
dimensions as the dynamics of a scalar field (the radion) coupled
with gravity in four dimensions. The radion
effective potential will enable us to discuss issues of
minimisation and stability in
section~\ref{radiondynamics:sect}~\footnote{See \cite{Carroll:2001ih}
  for a related discussion on the stability of these compactifications.}.
In particular we will show that only compactifications in
spherical spaces allow for a minimum of the radion potential with
positive or zero four dimensional cosmological constant, such as
the one observed. In case the compact space is hyperbolic, its
size can only be stabilised at the cost of a negative 4D
cosmological constant. The mass of the radion is of the order of
the compactification scale ($M_c$) in this scenario, as opposed to
radion masses of order $m_\phi \sim M_c^2/M_P$, with $M_P$ the
four-dimensional reduced Planck mass, used in other studies
of radion cosmology \cite{Kolb:2003mm}. This will change the
implications for cosmology of these extra-dimensional scenarios
with respect to models using other compactification mechanisms
that produce a suppressed radion mass \cite{Kolb:2003mm}. We will
also show that, even if the extra dimensions are compactified
in a spherical space, there is a maximum possible value for the
curvature of the 4D space above which the extra dimensions
decompactify, since for higher curvatures the radion effective
potential loses its minimum and exhibits a runaway
behaviour\footnote{See \cite{Giddings:2004vr} for a
  recent study of the decompactification process induced by thermal
  fluctuations or quantum tunnelling.}. This in particular puts a bound
on the maximum scale of inflation attainable in these models in terms
of the compactification scale: $V_{max} \sim M_c^2 M_P^2$.
We comment on other possible
contributions to the radion effective potential and how they
could modify the bounds obtained within flux compactifications.

\section{Compactifications with Fluxes}

In this section we review the spontaneous compactification of $n$
extra dimensions using fluxes and a higher dimensional cosmological
constant to
stabilise the size of the extra dimensions. If, starting in $d$
dimensions, we want to find static solutions that compactify $n$ of
them in a manifold of constant curvature, a natural thing to do is to
consider a vacuum expectation value for an $n$- or $4$-form field
strength (where $d=4+n$) \cite{Freund:1980xh}. Our starting point is
then the following
$d$-dimensional action (we follow Misner, Thorne and Wheeler's
book\cite{MTW:book} metric and curvature conventions):
\begin{equation}
S = \int d^d x \sqrt{-g}\left(M_\ast^{n+2}\frac{1}{2}R
-{1\over 2n!}F_{{\bf (n)}}^2 - {1\over
48}F_{{\bf (4)}}^2 - \hat{\Lambda}
\right)\label{lagrangian},
\end{equation}
where $M_\ast$ is the fundamental Planck mass and $\hat{\Lambda}$
is a higher-dimensional cosmological constant. We have just
considered pure gravity with a cosmological constant plus a
$n$-form and a $4$-form field. This Lagrangian can be seen as a
good approximation to study the dynamics of compactification if we
assume that all other fields present in our fundamental high
energy theory are stabilised with masses higher than those
relevant for our study (as we will see, the compactification
scale) or do not play any role in the compactification dynamics.
The corresponding equations of motion (EOM) resulting from this
action read:
\begin{equation}
\partial_M \left(\sqrt{-g} F_{\bf{(n)}}^{M..Q}\right)=0,
\end{equation}
\begin{equation}
\partial_M \left(\sqrt{-g} F_{\bf{(4)}}^{M..Q}\right)=0,
\end{equation}
and
\begin{eqnarray}
M_\ast^{n+2} R_{MN}  &= & T_{MN}-\frac{1}{n+2}g_{MN}T^R_{\quad R}
\nonumber \\
&=&
- g_{MN}{(n-1) \over n!(n+2)}F_{{\bf (n)}}^2
+  {1\over (n-1)!}{F_{\bf{(n)}}}_M^{\;\;\;P...Q}{F_{\bf{(n)}}}_{NP...Q}
\nonumber\\
&- &  g_{MN}{3 \over 24(n+2)}F_{{\bf (4)}}^2
+  {1\over
  6}{F_{\bf{(4)}}}_M^{\;\;\;P...Q}{F_{\bf{(4)}}}_{NP...Q}
+g_{MN}{2\over n+2} \hat{\Lambda}.
\end{eqnarray}

Although we require the existence of static solutions with $n$
compact dimensions, they are not the goal of our study. We will
focus on the dynamics of the volume of the extra dimensions. For
this we have to obtain the potential for the radion, $i.e.$ the
lower dimensional field that corresponds to dilatations of the
compact dimensions. For obtaining its EOM we will consider the
following metric \textit{ansatz}: 
\be ds^2 =
e^{-\alpha\hat{\phi}(x)}\gamma_{\mu \nu}(x)dx^\mu dx^\nu +
e^{-\beta\hat{\phi}(x)}R_0^2 \kappa_{ij}dz^i dz^j,
\label{metric:ansatz} 
\ee 
where latin indices run over the $n$
compact dimensions and greek ones run over the uncompactified
ones. $\kappa_{ij}$ is the metric for an Einstein manifold of
curvature $s=\pm 1$ (+1 like a sphere or -1 like an hyperbolic
plane) and $R_0$ is a constant with dimension $mass^{-1}$. We fix
the constants $\alpha=\sqrt{\frac{2n}{ n+2}}$ and
$\beta=-\frac{2}{n}\alpha$ in order to get a canonically
normalised kinetic term for $4$-dimensional gravity and for the
radion. The constant $R_0$ can be fixed to an arbitrary value
without loss of generality since, as can be seen from the metric, a
change in $R_0$ is equivalent to a shift in $\phi$ plus a
rescaling of the $x^{\mu}$ coordinates and the form fields.
Finally, we consider the following vacuum expectation value (VEV)
for the forms: \be F_{{\bf (n)}i...j} =\sqrt{\kappa}\hat{B}
\;\epsilon_{i...j}, \label{nFlux:ansatz} \ee
\begin{equation}
F_{(4)\;\mu\ldots \nu}=\sqrt{-\gamma}\hat{E}
\mathrm{e}^{-3\alpha \hat{\phi}} \epsilon_{\mu\ldots\nu},
\label{pFlux:ansatz}
\end{equation}
where $\hat{B}$, $\hat{E}$ are constants (of mass dimensions $(4-n)/2$ and
$(4+n)/2$, respectively), $\epsilon_{i...j}$ is
the totally antisymmetric tensor with $n$ indices and $\epsilon_{\mu...\nu}$ is
the totally antisymmetric tensor with 4 indices. The rest of the
elements of these forms are zero. Plugging this \textit{ansatz} in the
EOM we see that the equations of the forms are
satisfied and from the $(i,j)$ and $(\mu,\nu)$ components of Einstein
equations we get, respectively,
\begin{align}
M_\ast^{n+2}\Box_{(\gamma)} \hat{\phi}
=&
\sqrt{\frac{n}{2(n+2)}}
\Big[
-3\Big( \frac{\hat{B}^2}{R_0^{2n}}+\hat{E}^2
\Big)\mathrm{e}^{-3\alpha \hat{\phi}}
-2 \Lambda \mathrm{e}^{-\alpha \phi}
+\frac{n+2}{n}s\frac{M_\ast^{n+2}}{R_0^2}
\mathrm{e}^{-\frac{n+2}{n}\alpha\hat{\phi}}
\Big],
\label{EoM6D4D:radion}
\end{align}
and
\begin{align}
M_\ast^{n+2}
R_{\mu\nu}(\gamma)=&
M_\ast^{n+2}
\partial_\mu \hat{\phi} \partial_\nu \hat{\phi}
+
\frac{1}{2}\Big[ \frac{\hat{B}^2}{R_0^{2n}}+\hat{E}^2 \Big]
\mathrm{e}^{-3\alpha \hat{\phi}} \gamma_{\mu\nu}
\nonumber \\
&+\hat{\Lambda}\mathrm{e}^{-\alpha\hat{\phi}} \gamma_{\mu\nu}
-
\frac{1}{2}
s\frac{M_\ast^{n+2}}{R_0^2}
\mathrm{e}^{-\frac{n+2}{n}\alpha \hat{\phi}}
\gamma_{\mu\nu},
\label{EoM6D4D:gravity}
\end{align}
where $\Box_{(\gamma)}$ and $R_{\mu\nu}(\gamma)$ are the d'Alembertian
and the Ricci tensor computed with the metric $\gamma_{\mu \nu}$. One
can check that these EOM can be derived also varying
with respect to $\gamma_{\mu \nu}$ and $\hat{\phi}$ the four-dimensional
action
\begin{eqnarray}
 S_\mathrm{eq}
&=&
 V_n
 \int \mathrm{d}^4 x \;
 \sqrt{-\gamma}\;
 \Big[
 \frac{1}{2}M_\ast^{n+2} R_{(\gamma)} -
\frac{1}{2}M_\ast^{n+2}\partial_\mu \hat{\phi}
 \partial^\mu \hat{\phi}
  \nonumber\\
&&
\phantom{
V_n
 \int \mathrm{d}^p x \;
}
-\frac{1}{2}\Big[ \frac{\hat{B}^2}{R_0^{2n}}
+
\hat{E}^2 \Big] \mathrm{e}^{-3\alpha\hat{\phi}}
- \hat{\Lambda} \mathrm{e}^{-\alpha\hat{\phi}}
+\frac{1}{2}s\frac{M_\ast^{n+2}}{R_0^2}e^{-\frac{n+2}{n}\alpha\hat{\phi}}
\Big]
\nonumber \\
&=&
 \int \mathrm{d}^4 x \;
 \sqrt{-\gamma}\;
 \Big[
 \frac{1}{2}M_P^{2} R_{(\gamma)} -
\frac{1}{2}\partial_\mu \phi
 \partial^\mu \phi
  \nonumber\\
&&
\phantom{
V_n
 \int \mathrm{d}^p x \;
}
-
E^2
\mathrm{e}^{-3\alpha\phi/M_P}
- \Lambda \mathrm{e}^{-\alpha\phi/M_P}
+\frac{1}{2}s\frac{M_P^{2}}{R_0^2}e^{-\frac{n+2}{n}\alpha\phi/M_P}
\Big],
\label{effaction}
\end{eqnarray}
where the volume of the internal dimension is $V_n = \int d^nz\,\sqrt{k}
R_0^n$ and in the last two lines we have canonically normalised
the action by defining the four-dimensional Planck mass as
$M_P^2=M_\ast^{n+2}V_n$ and redefining $\hat{\phi}=\phi/M_P$. We
have also defined the corresponding ``effective four-dimensional
flux''
\begin{equation}
E^2=\frac{V_n}{2}\Big[\frac{\hat{B}^2}{R_0^{2n}}+\hat{E}^2 \Big],
\end{equation}
and a four-dimensional cosmological constant $\Lambda=\hat{\Lambda}V_n$.
An important issue in these compactifications is the volume of the
extra dimensional manifold, since it enters the definition of all
four-dimensional parameters in terms of the higher-dimensional ones.
In the hyperbolic case we can construct compact manifolds of constant
curvature by modding out the (non-compact) universal covering space of
$n$-dimensional hyperbolic manifolds by certain discrete subgroups of
its
isometry group \cite{wolf}. Similarly, in the spherical case, we can
simply consider the $n$-dimensional sphere (since it is already
compact) or we can consider other non-trivial topologies by modding
out by some discrete subgroup of its
isometry group. In this case the volume of the space constructed in such a
way will be $\frac{vol(S^n)}{|\Gamma|}$, where $|\Gamma|$ is the
number of elements of the discrete group. For even-dimensional
manifolds we only have two possibilities, a sphere and the projective
space obtained from the sphere identifying antipodal points, being the
volume of the former twice that of the latter. For odd-dimensional
spaces we have more possibilities, and in particular we can use the
cyclic group $Z_q$ of arbitrary order. For this reason we will leave
the volume of the compactification manifold as a free parameter, not
related directly with the curvature, but keeping in mind that in the
spherical case the maximum possible volume will be the volume of the
$n$-sphere.

Another important point to remark is that the effective 4D action,
eq.(\ref{effaction}), is $not$ the original $d$-dimensional action
with the \textit{ansatze} given by
eqs.(\ref{metric:ansatz}-\ref{pFlux:ansatz}) substituted in it.
There is a difference in the sign corresponding to the $4$-form
term $\Delta {\cal L}=-\frac{1}{2}V_{n} \hat{E}^2
\mathrm{e}^{-3\alpha\phi}$, since substituting the
\textit{ansatze} eq.(\ref{metric:ansatz}-\ref{pFlux:ansatz})
naively in the action would have resulted in a contribution to
this effective action given by $\Delta {\cal L}=\frac{1}{2}V_{n}
\hat{E}^2 \mathrm{e}^{-3\alpha\phi}$, that gives the wrong
EOM (see \cite{Duff:1989ah}). The reason for
this is the following: if we substitute in an action the VEV of
the derivative of a field in terms of other fields (as we have
made for the forms, since $F_{\bf{4}}=d
A_{\bf{3}}=\frac{E}{\sqrt{-\gamma}} \mathrm{e}^{-3\alpha \phi(x)}
\epsilon_{\mu\ldots\nu}$) and we vary the action with respect to
these fields ($\phi(x)$ in our case) we are not going to obtain
the correct EOM, since in the original action we
were varying with respect to $A_{\bf{3}}$, and given that this
field appears with a derivative in that term, and $\phi$ does not,
such a procedure is not justified\footnote{We thank
  J.A. Casas for clarifying this point to us.}. The action
(\ref{effaction}), although does $not$ come from the higher
dimensional one substituting the particular \textit{ansatze} we have
chosen, does produce the correct EOM, completely equivalent to the
higher-dimensional ones.

\section{Radion Dynamics and Inflation\label{radiondynamics:sect}}

\begin{figure}[h]
  \begin{center}
    \epsfig{file=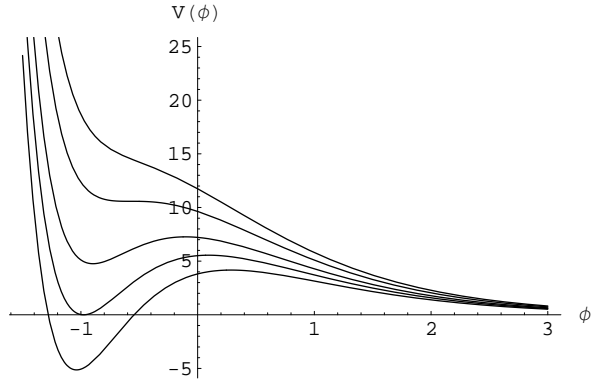,height=5cm}
  \end{center}
  \caption{Radion potential for different values of $\Lambda$ with
  fixed $E$ and $R_0$. The different lines are for $\Lambda$ equal to,
from top to bottom, $1.1 \Lambda_\mathrm{max}$,
  $\Lambda_\mathrm{max}$, $1.1 \Lambda_0$, $\Lambda_0$ and $0.9
  \Lambda_0$, corresponding to unstable and marginally stable extra
  dimensions the first two and stable extra dimensions with positive,
  zero and negative effective four-dimensional cosmological constant
  the lower three. \label{radionpotential:plot}}
\end{figure}

As we have shown in the previous section, the dynamics of the
volume modulus in flux compactifications is reduced to the
dynamics of a scalar field in curved space with the following
potential
\begin{equation}
V(\phi)= E^2 e^{-3\alpha \phi/M_P}
+\Lambda e^{-\alpha \phi/M_P}
-\frac{1}{2}s\frac{M_P^2}{R_0^2}  e^{-\frac{n+2}{n}\alpha\phi/M_P}.
\label{potential}
\end{equation}
Several interesting points can be highlighted by inspecting this
potential. First, as was discussed in ref.\cite{Giddings:2004vr},
$V\rightarrow 0$ for $\phi \rightarrow \infty $, so in case the
potential has a minimum with $V>0$ it is necessarily
metastable\footnote{In \cite{Giddings:2004vr} it was
  argued that this property applies to a wider class of
  compactifications, using not only fluxes in order to generate the
  radion potential, but also branes and non-perturbative
  effects.}, but in any case the lifetime of the decay through quantum
tunneling can be easily made bigger that the age of the
universe. However, it is not guaranteed that there is a minimum at 
finite $\phi$. For this to happen the following conditions have to be
satisfied
\begin{equation}
\Lambda \leq \Lambda_\mathrm{max}\equiv
3(n-1)E^2 \bigg(\frac{n+2}{6n^2}\frac{M_P^2}{E^2R_0^2}
\bigg)^{\frac{n}{n-1}},
\label{Lambdamax:sphere}
\end{equation}
for $s=+1$ and
\begin{equation}
\Lambda \leq 0,
\label{Lambdamax:hyperbolic}
\end{equation}
for $s=-1$.
Assuming the corresponding condition is fulfilled, \textit{i.e.} there is a
finite value $\phi_0$ such that $V^\prime(\phi_0)=0$, we have, for the
effective four-dimensional cosmological constant
\begin{equation}
V(\phi_0)=-2 E^2 e^{-3\alpha \phi_0/M_P} + \frac{s}{n}\frac{M_P^2}{R_0^2}
e^{-\frac{n+2}{n}\alpha \phi_0/M_P},\label{Vphi0}
\end{equation}
which immediately shows that compactification on a hyperbolic
manifold ($s=-1$) leads to a negative cosmological constant in the
lower dimensional theory. Compactifications on
positively curved manifolds can on the other hand result on a
four-dimensional cosmological constant of any sign, depending on
the value of the higher-dimensional one. If it is tuned against
the fluxes as
\begin{equation}
\Lambda=\Lambda_0\equiv(n-1)E^2
\bigg(\frac{M_P^2}{2nE^2R_0^2}\bigg)^{\frac{n}{n-1}},
\label{Lambda:0}
\end{equation}
the four-dimensional cosmological
constant vanishes, being positive or negative
when $\Lambda$ is larger or smaller than $\Lambda_0$,
respectively. 
The above conclusions strongly rely on the fact that the flux
contribution to the radion potential has a well defined sign. It is
possible to obtain new contributions to the radion potential with the
opposite sign, for instance those coming from a wrapped $p-$brane (with
$3\leq p\leq 2+n$ since the contribution of a $n+3$ brane is
equivalent to that of $\hat{\Lambda}$) that would give a contribution
to the radion potential proportional to its tension
(see for instance \cite{Giddings:2004vr}),
\begin{equation}
V_p(\phi) = T_p e^{-\frac{2n+3-p}{n} \alpha \phi/M_P},
\label{brane:pot}
\end{equation}
where we have conveniently absorbed in the definition of the brane
tension, $T_p$, the the volume and other numerical factors appearing
in the dimensional reduction. It is of course not guaranteed that the
introduction of those branes will not modify the background in an
important way. In fact, one would expect that the internal manifold
would no longer be Einstein but acquire warping. Furthermore,
singularities will in general appear in the classical description of
the background~\cite{Horowitz:cd}. However, 
for the sake of the present
argument we will assume that the backreaction of the branes
can be neglected so that their only effect is to add a
contribution to the radion potential like the one in
eq.(\ref{brane:pot}). 
The exponent shows that this term is
irrelevant for $\phi \to \pm \infty$, provided fluxes and a
higher-dimensional cosmological constant are present. Its generic
effect is therefore to increase or decrease the potential at
intermediate values of $\phi$ for positive and negative tensions,
respectively. This means that positive tension branes can be used to
increase the effective four-dimensional cosmological constant (see
\cite{Kachru:2003aw} for a recent application to uplift AdS to dS
vacuum in string theory).
Nonetheless one should be cautious when doing this
since too large a positive contribution can ruin
the existence of a minimum for the radion potential and therefore
destabilise the extra dimensions (see also~\cite{Burgess:2003ic}).
Negative tension branes on the other hand,
tend to decrease the effective four-dimensional
cosmological constant but they also tend to stabilise the radion
potential. In particular it is possible to find values of the
parameters such that the radion is stabilised with hyperbolic compact
dimensions and a positive four-dimensional cosmological constant but
only at the cost of introducing these branes with \textit{negative}
tension, (the potential would be unstable without the branes, that
decrease it in the intermediate region so that it develops a minimum
before the value of the potential crosses zero), as can be seen from
the value of the potential at the minimum including the effect of the
branes
\begin{equation}
V(\phi_0)=-2 E^2 e^{-3\alpha \phi_0/M_P} - \frac{1}{n}\frac{M_P^2}{R_0^2}
e^{-\frac{n+2}{n}\alpha \phi_0/M_P}-\frac{n+3-p}{n}T_p
e^{-\frac{2n+3-p}{n} \alpha \phi_0/M_P}.
\end{equation}
Given the fact that negative tension branes usually suffer
from severe stability problems and the possible important effects of
the backreaction of the branes on the background, 
we prefer not to pursue this
possibility any further, but restrict ourselves to the potential in
eq.(\ref{potential}) arising exclusively from fluxes and a
higher-dimensional cosmological constant.
(A review of the recent attemps to obtain moduli stabilization in
string theory can be found in \cite{Silverstein:2004id}.)
In that case, the phenomenological restriction of having a positive (and
eventually small) four-dimensional cosmological constant prevents the
use of hyperbolic compactifications that we will not consider any more.
 All the
qualitatively different behaviours for the radion that can be obtained
in flux compactifications on a positively curved manifold are
graphically summarised
in fig.~\ref{radionpotential:plot}, where
we have displayed the radion potential
for fixed values of $E$ and $R_0$ and the
following values of $\Lambda$,
$\Lambda=0.9 \Lambda_0, \Lambda_0, 1.1 \Lambda_0,
\Lambda_\mathrm{max}$ and $1.1 \Lambda_\mathrm{max}$, corresponding to
stable extra dimensions with negative, zero and positive effective
four-dimensional cosmological constant, respectively the first three,
and marginally stable and unstable the last two.
In particular, using the fact that $0 \leq \Lambda_0 \leq
\Lambda_\mathrm{max}$, we see how, starting with the radion on a
stable or metastable minimum, any contribution that increases the
higher-dimensional cosmological constant can destabilise the
system leading to a run-away potential for the radion as in the
top plot in the figure. In that case the extra dimensions would
decompactify, rendering the model unrealistic. This possibility
will allow us to put stringent bounds on the scale of inflation,
using the fact that the couplings of the radion are completely
determined by general covariance\footnote{It can be argued that
radiative corrections arising below the compactification scale
could spoil the form of the radion potential that ultimately comes
from 6D general covariance. However, even if these corrections are
present one would expect them to be of order $\delta V \sim M_c^4
f_1(\phi/M_P)$, that represent subdominant corrections with
respect to the terms we are already considering in the potential
that are of order $V\sim M_c^2 M_P^2 f_2(\phi/M_P)$, where $f_1$
and $f_2$ are dimensionless functions.}.

In order to obtain a realistic situation we consider that the
higher-dimensional cosmological constant is tuned to give a vanishing
four-dimensional cosmological constant, $\Lambda=\Lambda_0$.
The compactification scale can then be written as
\begin{equation}
M_c^2 = \frac{1}{R_0^2} e^{-\frac{n+2}{n} \alpha
  \phi_0}=\frac{1}{R_0^2}
\bigg(\frac{M_P^2}{2nE^2R_0^2} \bigg)^\frac{n+2}{2(n-1)},
\end{equation}
the different Kaluza-Klein modes appearing in our theory will have
masses proportional to this compactification scale. The precise value
depends on the topology of our compact space, so they could be
significantly higher. We can use this expression of the
compactification
scale to trade $R_0$ for it in all our formulae, writing everything
as a function of $E$, $M_P$ and $M_c$. For instance, our tuned
cosmological constant reads
\begin{equation}
\Lambda_0=(n-1)\bigg(\frac{M_P^2 M_c^2 E}{2n}\bigg)^{2/3},
\end{equation}
and similarly for the critical value of the cosmological constant for
a stabilised radion that can be written
\begin{equation}
\Lambda_\mathrm{max}=3(n-1)
\bigg(\frac{n+2}{3n}\bigg)^\frac{n}{n-1}
\bigg(\frac{M_P^2 M_c^2 E}{2n}\bigg)^{2/3}.
\end{equation}
The radion mass, given by the second derivative of its potential at
the minimum, turns out to be of the order of the compactification scale,
\begin{equation}
m_\phi^2=V^{\prime\prime}(\phi_0)=4 \frac{n-1}{n(n+2)} M_c^2.
\end{equation}
This mass is much larger than the normally used estimates that
suppress it by the ratio $M_c/M_P$. This last estimate is usually
correct when the extra dimensions are flat and the radion is
massless at leading order (see for instance
\cite{Albrecht:2001cp}). In our case, however, the fluxes and
higher-dimensional cosmological constant compactify the extra
dimensions in a highly curved space and therefore give a mass to
the radion that is much larger than the naive
estimate\footnote{This does not mean that in these kind of
  compactifications there could not be some light fields with mass $m
  \sim M_c^2/M_P$. For instance in the SUGRA of
  ref.\cite{Nishino:1984gk} two extra dimensions can be compactified
  in a sphere using a magnetic field but there is a linear combination
  of the radion and dilaton that remains massless after
  compactification. It has been argued that a suppressed mass for this
  field is generated after SUSY is broken \cite{Burgess:2004yq}. We
  are assuming that no such light field is present here.}. This is
very interesting because a Planck suppressed
mass for the radion typically produces very stringent cosmological
constraints~\cite{Kolb:2003mm},
constraints that are easily eluded in our case.
Using this value of the radion mass and the fact that its couplings
are Planck suppressed, we can estimate its decay width as
\begin{equation}
\Gamma_\phi=\tau_\phi^{-1} \sim \frac{m_\phi^3}{M_P^2}
\sim \frac{M_c^3}{M_P^2}.
\end{equation}
This indicates that the radion decays before BBN for $M_c \gtrsim 10$
TeV, decays after BBN for $10$ MeV $\lesssim M_c \lesssim 10$ TeV and
is effectively stable for $M_c \lesssim 10$ MeV. In the last two
cases, constraints on the compactification scale could arise
from modifications of the successful predictions of BBN or the CMB
spectrum (see \cite{Holtmann:1998gd}) and
over-closure of the universe, respectively. A more detailed study of
such possibilities is deferred to future work.

The last quantity relevant for our discussion is the value of the
potential at the critical point, where it 
has a saddle point instead of a minimum when
$\Lambda=\Lambda_\mathrm{max}$, (second from the top in
fig.~\ref{radionpotential:plot}),
\begin{equation}
V_{max}=V(\phi_\mathrm{max},\Lambda_\mathrm{max})=
2 \frac{n-1}{n(n+2)} \bigg( \frac{n+2}{3n}\bigg)^\frac{3n}{2(n-1)}
M_P^2 M_c^2,\label{Vmax}
\end{equation}
where we have denoted by $\phi_\mathrm{max}$ the value of the
radion such that $V^\prime(\phi_\mathrm{max})=
V^{\prime\prime}(\phi_\mathrm{max})=0$ for
$\Lambda=\Lambda_\mathrm{max}$. Although the scale of the 4D
effective potential at this critical point ($V_{max}^{1/4}$) is
above the compactification scale (and can be even above the higher
dimensional Planck mass $M_{\ast}$), the use of a field theoretic
4D description is fully justified. The destabilisation occurs when
the curvature of the compact dimensions and the four non-compact
ones are roughly of the same order, $\sim M_c^2$, that we assume
to be well below the higher dimensional fundamental mass (so that
higher order curvature corrections are negligible). And for
the 4D description to be a good approximation we just have to make
sure that the temperature is always below the compactification
scale, so KK excitations will not be produced.

We are now in a position to discuss the phenomenological
implications of flux compactifications on positively curved
internal manifolds for inflation. We consider that the
higher-dimensional cosmological constant has been tuned to give a
vanishing effective four-dimensional constant at the minimum of
the potential, $\Lambda=\Lambda_0$. The first possibility is to
consider the radion as the inflaton itself. Of course a detailed
study is necessary in order to determine the viability of the
model although in principle it seems plausible that enough number
of e-foldings can be obtained by appropriately tuning the initial
position of the radion as close as necessary to its maximum. In
this case it is clear that we can not obtain inflation scales
higher than the maximum of the $\Lambda = \Lambda_0$ curve in
fig.1 (the one with the minimum in $V=0$).

The other obvious possibility is considering that inflation is
driven by \textit{any} other scalar field but the radion. In this
case we can get higher inflation scales, but when getting bounds
on this scale the interesting thing is that it does not really
matter the details of inflation.  What matters is the fact that
during inflation the slow-roll condition ensures that the inflaton
potential is almost constant so the only effect on the radion
potential is just to add a contribution to the higher-dimensional
cosmological constant equal to the inflaton potential:
\begin{equation}
\Lambda \to \Lambda+V(\chi),
\end{equation}
where $V(\chi)$ is the inflaton potential, that can be considered
constant as long as the slow-roll condition (along the $\chi$
direction) holds. It is then evident from our previous discussion
that for
\begin{equation}
V(\chi) \geq \Lambda_\mathrm{max}-\Lambda_0,
\end{equation}
the radion potential loses its minimum and the extra dimensions
decompactify. This inequality is precise only when the slow-roll
condition for the inflaton potential is maintained but can be
considered as a reasonable order of magnitude estimate for
realistic situations. Of course, the details of the inflaton
potential and the initial conditions are required for a detailed
computation of the scale at which our theory decompactifies.
However, the fact that the very slow-roll condition can be
jeopardised if the inflaton potential is higher that this
maximum scale (since the slope would be much larger than expected
in the radion direction of the two-dimensional potential),
indicates that eq.(\ref{Vmax}) is a conservative estimate for the
maximum scale of inflation, where we have assumed that the extra
dimensions stabilise before inflation takes place. 
This bound can have important consequences for many models of
inflation in extra-dimensional theories. For instance, in the recent
study of 
inflation in 6D gauged supergravity~\cite{Brandenberger:2004vz} the
authors consider the possibility of chaotic inflation in an anomaly 
free $N=1$ gauged supergravity in six
dimensions~\cite{Randjbar-Daemi:wc}, compactified with fluxes on a
two-sphere. Their chaotic potential is, during inflation, of order
\begin{equation}
V_{ch} \sim M_c^2 M_P^2,
\end{equation}
which is on the verge of the decompactification limit. This is a clear example in which the possibility of
destabilization of the extra dimensions during inflation has to be
carefully taken into account in order to determine the phenomenological
viability of the model.

The bound we have found on the maximum scale of inflation seems to
disfavour low scale compactifications, since the CMB data prefers high inflation scales:
\begin{equation}
\Big(\frac{V_{infl}}{\epsilon}\Big)^{1/4} \sim 10^{16}GeV,
\end{equation}
where $\epsilon = M_P^2 (V^{\prime}/V)^2$
is the slow roll parameter and this is valid only when the
inflaton is responsible for the density perturbations
(see \cite{Trodden:2004st} and references therein). 
As an
example of the bounds implied, for $n=2$ extra dimensions of size
$M_c\sim 10^{-3}$ eV, the maximum allowed scale of inflation is
$\sim\, \mathrm{TeV}^4$, whereas for TeV-sized extra dimensions we
obtain $V_\mathrm{max}\sim (10^{10}-10^{11}\, \mathrm{GeV})^4$.

Another effect that could be important when describing the stability
of flux compactifications is finite temperature corrections. It has
been very recently shown in \cite{Buchmuller:2004xr} that, due to the
fact that gauge couplings are dilaton dependent, finite
temperature effects $\delta V \propto g^2 T^4$, 
can destabilise the dilaton potential for temperatures
above $10^{11}-10^{12}$ GeV. 
A similar effect can also affect the
radion potential since, as can be easily seen from the kinetic term of
a bulk gauge boson, in our case the gauge couplings are radion
dependent
\begin{equation}
g^2=\frac{g_d^2}{V_n} e^{-\alpha\, \phi/M_P},
\end{equation}
where $g_d$ is the higher-dimensional gauge coupling. 
Therefore we see that the finite temperature contribution to the
radion potential scales as the cosmological constant term and thus
this effect can be taken into account by simply replacing
\begin{equation}
\Lambda \to \Lambda+ \xi T^4,
\end{equation}
where the parameter $\xi$ is a number of order one that depends on the
number of interacting species. This leads to a maximum temperature in
the early universe of the order of
\begin{equation}
T_\mathrm{max}^4 \sim \Lambda_\mathrm{max} - \Lambda_0 \sim (M_c^2
M_P^2 E)^{2/3}\label{boundT}.
\end{equation}
Note that, unless $E\leq M_c^4/M_P^2$, this temperature is much
larger than the compactification scale and therefore, beyond the
validity of our four-dimensional approximation. A full
higher-dimensional study should then be performed to give precise
bounds on the maximum attainable temperature from
decompactification. Nonetheless, naively one would expect that the higher
number of active species when the Kaluza-Klein modes are excited would
increase the finite temperature correction to the radion effective
potential, so the bound given by eq.(\ref{boundT}) could be considered
a conservative one.  

As a final remark we would like to mention that a similar bound as
the one we have placed on the maximum scale of inflation in flux
compactifications applies to a much more general range of
compactifications. It has been recently argued
in~\cite{Giddings:2004vr} that not only fluxes or a higher
dimensional cosmological constant but also space-filling branes,
non-perturbative effects or higher order string corrections lead
in string theory to a qualitatively similar potential to the one
we have depicted in fig.~\ref{radionpotential:plot}. The effect of
branes has already been discussed in the previous section.
Non-perturbative effects are argued in~\cite{Kachru:2003aw} to
give rise to the following potential
\begin{equation}
\delta V_{NP} \sim B_{NP} e^{-2a e^{\frac{2}{3} \alpha \phi/M_P}}
e^{-\frac{2}{3} \alpha
  \phi/M_P},
\end{equation}
where $a$ is here a parameter that depends on the mechanism to
generate such corrections, whereas higher order string corrections
give a contribution~\cite{Becker:2002nn}
\begin{equation}
\delta V_{HO} \sim B_{\alpha^\prime} e^{-3 \alpha\phi/M_P}.
\end{equation}
Non-perturbative effects are again irrelevant for $\phi \to \pm
\infty$ so their effect is similar to the one of $p-$branes in the
sense that they only modify the intermediate regions of the potential,
while the effect of higher order string corrections is similar to that
of the fluxes and can be accounted for by simply shifting the constant
$E^2$ in our analysis.
In any case, the relevant point is that
these corrections do not modify in a qualitative way the radion
potential.
Thus, a positive contribution to the
cosmological constant (like a slow-rolling inflaton)
will tend to remove the (meta)stable minimum for the radion
potential and therefore a bound on the maximum curvature
during inflation similar to the one we have discussed can be put,
although the details will of course depend on the sources for radion
stabilisation used.

\section{Conclusions}

In this letter we have studied the dynamics of the size of the extra
dimensions (the radion) in flux compactifications. We have obtained
the radion effective potential and extracted some interesting results
from it. First, we have shown that the requirement that there is a
minimum with positive or zero 4D cosmological constant implies that
the compactification manifold has spherical curvature, hyperbolic
compactifications being ruled out if fluxes are responsible for the
stabilisation of the extra-dimensional volume. Then, we have seen that
there is a maximum value for the de Sitter curvature of the 4D space
for which the radion can be stabilised, and this sets an upper bound
on the scale of inflation from decompactification of the extra
dimensions. This maximum value occurs when the 4D space and the extra
dimensions have roughly the same curvature $\sim M_c^2$, so the limit
on the scale of inflation is of order $V_{max} \sim M_c^2 M_p^2$. For
the extreme ADD case with $M_c \sim 10^{-3}$ eV and $n=2$ the maximum
scale of inflation is ${\cal O}(\mathrm{TeV})$, while for TeV size extra
dimensions this bound implies that the maximum scale is ${\cal
  O}(10^{10}-10^{11}\mathrm{GeV})$.
This bound seems to disfavour models with a
low compactification scale, since CMB data prefers high inflation
scales (at least when the inflaton is responsible for the density
perturbations).
Finally, we have briefly mentioned other possible contributions to the
radion potential. Finite temperature effects, that can act as an
effective cosmological constant and therefore destabilise the radion
have been shown to be negligible within the range of validity of our
effective four-dimensional description.
Other sources that may appear in string theory, like
wrapped $p-$branes, non-perturbative effects of higher order string
corrections do not qualitatively change the picture, still existing a
maximum value of the inflation scale above which the extra dimensions
would decompactify, although the details of such scale depend on the
particular contributions.

\section*{Acknowledgements}

It is a pleasure to thank J.A. Casas and S. Davidson for
collaboration at the early stages of this work and very fruitful
discussions. We also want to thank S.A. Abel and C.P. Burgess
for useful comments. This work has been funded by PPARC.


\end{document}